\documentclass[prl,twocolumn,showpacs,floatfix]{revtex4}
\usepackage{times}

\usepackage{graphicx}
  \graphicspath{{./Eps/}}
\usepackage{amsmath}
\usepackage{amssymb}

\newcommand{\vc}[1]{
\textbf{\textit{#1}} }

\begin{document}

\title{Ballistic persistent currents in disordered metallic rings: Origin of puzzling experimental values}

\author{J. \surname{Feilhauer}}

\author{M. \surname{Mo\v{s}ko}}
\email{martin.mosko@savba.sk}

\affiliation{Institute of Electrical Engineering, Slovak
Academy of Sciences, 841 04 Bratislava, Slovakia}

\date{\today}

\begin{abstract}
Typical persistent current ($I_{typ}$) in a normal metal ring with disorder due to random grain boundaries and rough edges is calculated microscopically. If disorder is due to the rough edges, a ballistic current $I_{typ}\simeq e v_F/L$ is found in spite of the diffusive resistance  ($ \propto L/l$), where $v_F$
is the Fermi velocity, $l$ is the mean free path, and $L \gg l$ is the ring length. This ballistic current has a simple interpretation: It is due to a single electron that moves (almost) in parallel with the rough edges and thus hits them rarely. Our finding agrees with
a puzzling experimental result $I_{typ}\simeq e v_F/L$, reported by Chandrasekhar et al. [Phys. Rev. Lett. \textbf{67}, 3578 (1991)] for metal rings of length $L \simeq 100 l$. If disorder is due to the grain boundaries, our results agree with theoretical result $I_{typ}\simeq (e v_F/L) (l/L)$ that holds for the white-noise-like disorder and has been observed in recent experiments.
\end{abstract}

\pacs{73.23.-b, 73.23.Ra}
\keywords{quasi one-dimensional transport, surface roughness, quantum conductance,
universal conductance fluctuations}

\maketitle


A mesoscopic resistive metal ring pierced by magnetic flux ($\Phi$) supports a
persistent
current \cite{Buttiker,Chand,Jariwala,Bluhm,Bles}.
At zero temperature, the ring supports the persistent current
$I = {\sum}_{ \forall E_j \leq E_F} I_j$, where $I_j(\Phi) = -dE_j(\Phi)/d\Phi$ is the single-electron current carried by electron with eigen-energy $E_j(\Phi)$, and $E_F$ is the Fermi level \cite{Buttiker}. Function $I(\Phi)$ is periodic with
period $\Phi_0 \equiv h/e$, which provides a clear-cut experimental sign of the persistent current \cite{Buttiker,Chand,Jariwala,Bluhm,Bles}.
If the ring is ballistic and possesses one conducting channel, the sum $\sum I_j$ changes its sign when a new occupied state $j$ is added.
Due to the sign cancelation mainly the electron at the Fermi level contributes to the sum, and the amplitude of the current is $I_0 = ev_F/L$,
where $v_F$ is the Fermi velocity and $L$ is the ring circumference.
If the ring is disordered, the size and sign of
the current fluctuate from sample to sample due to the disorder fluctuations.
It is then reasonable to asses a typical current in a single sample as $I_{typ}=\langle I^2\rangle ^{1/2}$, where $\langle \dots \rangle$ means the ensemble average.

The number of the conducting channels ($N_c$) in disordered metal rings is typically $\gg 1$ and the rings obey the diffusive limit, $l \ll L \ll \xi$, where $l$ is the electron mean free path and $\xi \simeq N_c l$ is the localization length.
 To estimate $I_{typ}$, one can assume again that mainly the electron at the Fermi level contributes to the sum $\sum I_j$. Since $L \gg l$, the electron is expected to move around the ring by diffusion. Its transit time is $\tau_D = L^2/D$, where $D = v_F l/d$ is the diffusion coefficient and $d$ is the dimensionality of the sample. So $I_{typ} \simeq e/\tau_D = (1/d)(e v_F/L)(l/L)$.
  A similar result follows from the Green functions theory for non-interacting electrons \cite{Cheung}, if disorder is modeled by a random potential $V(\bold{r})$ obeying the white-noise condition $\langle V(\bold{r}) V(\bold{r}')\rangle \propto \delta(\bold{r} - \bold{r}')$. One obtains \cite{Cheung}
\vspace{-0.6cm}
\begin{equation}
\vspace{-0.1cm}
I^{theor}_{typ} = 2 \times (1.6/d) (e v_F/L) (l/L), \quad l \ll L \ll \xi .
 \label{Igre}
\end{equation}
Here $2$ is the spin factor, $d =1$, $2$, $3$, and $1.6$ is from Ref. \cite{Feilhauer1}.

The first observation of persistent current in a single metallic ring was reported \cite{Chand} for three Au rings of size $L \sim 100 l$. The
measured currents
were ten-to-hundred times larger than the result \eqref{Igre}; they ranged from $\sim 0.1ev_F/L$ to $\sim ev_F/L$.
This huge discrepancy has not been explained yet (see the reviews \cite{Saminadayar,Shanks}).
Other Au rings showed \cite{Jariwala} currents a few times larger than result \eqref{Igre}, and recent measurements of individual Au rings \cite{Bluhm} and Al rings \cite{Bles} agreed with result \eqref{Igre} very well.

Why did the similar measurements of diffusive Au rings \cite{Chand,Bluhm} show quite different results, $I_{typ}\simeq e v_F/L$ and
$I_{typ}\simeq (e v_F/L) (l/L)$? A puzzle \cite{Chand} is why a multichannel disordered ring of length $L \gg l$ carries the current $\sim e v_F/L$, typical for a one-channel ballistic ring? This Letter wants to answer both questions. There is disorder due to polycrystalline grains and rough edges \cite{Saminadayar}
even in pure Au rings. Using a scattering-matrix method for non-interacting electrons \cite{Feilhauer1,Feilhauer2}, we study typical persistent currents in Au rings with grains and rough edges \emph{without the white-noise approximation}.

If the disorder is due to the grains, our results agree with the white-noise-related formula \eqref{Igre} and experiments \cite{Bluhm,Bles}. However, if the disorder is due to the rough edges, we find the  ballistic-like result $I_{typ}\simeq e v_F/L$ albeit the resistance is diffusive ($\propto L/l$) and $L \gg l$, like in the experiment \cite{Chand}. This ballistic current is due to a single electron that moves (almost) in parallel with the rough edges and thus hits them rarely. Briefly,
result $I_{typ}\simeq e v_F/L$ in a metal ring of length $L \gg l$ is as normal as the
result $I_{typ}\simeq (e v_F/L)(l/L)$. Which result is observed depends on the nature of disorder.

\begin{figure}[t!]
\centerline{\includegraphics[clip,width=0.9\linewidth]{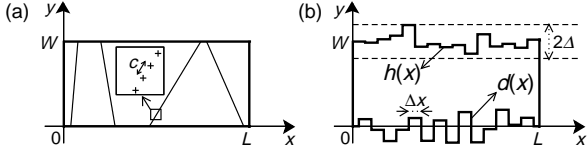}}
\caption{Model \cite{Feilhauer1,Feilhauer2} of wires with grain boundaries (a) and rough edges (b). Let $d(x)$ and $h(x)$ be the $y$-coordinates of the edges. Then $V(x,y)=0$ for $d(x)<y<h(x)$ and
$V(x,y)=\infty$ elsewhere. For smooth edges $d(x)=0$ and $h(x)=W$, otherwise $d(x)$ and $h(x)$ fluctuate randomly in intervals $\langle -\Delta, \Delta \rangle$ and $\langle W-\Delta, W+\Delta \rangle$. The RMS of the fluctuations is $\delta = \Delta/\sqrt{3}$. The fluctuations occur abruptly with step $\Delta x$. So $\Delta x$ is the roughness correlation length.
 The grain boundaries (a) are modeled as a randomly-oriented mutually non-intersecting lines, that is, the angle between the given line and the $x$ coordinate is random in a range restricted by the presence of the neighboring lines \cite{Feilhauer2}. Each line  consists of equidistant repulsive dots (plus signs) with potentials $\gamma \delta(x-x_i) \delta(y - y_i)$, where ($x_i$, $y_i$) is the position of the $i$-th dot. Thus
$U(x,y) = {\sum}_{ \forall i} \gamma \delta(x - x_i) \delta(y - y_i)$. If $c \to 0$ and $\gamma/c$ is fixed, a boundary scatters electrons as a structure-less
 line-shaped barrier independent on the choice of $c$. If a 2D electron impinges on such a barrier perpendicularly with Fermi wave vector $k_F$, it is reflected with probability
$R_G = (\bar{\gamma}/c)^2/[k^2_F + (\bar{\gamma}/c)^2]$,
where $\bar{\gamma} = m \gamma / \hbar^2$. The parameters of the grain boundary disorder are $R_G$ (in reality \cite{comment00} $R_G \sim 0.1 - 0.8$) and the mean inter-boundary distance $d_G$.} \label{Fig-1}
\vspace{-0.5cm}
\end{figure}

 For simplicity, we study two-dimensional (2D) rings and mention the 3D effect at the end. We start with a conductance study. Consider a 2D wire [Fig. \ref{Fig-1}] described by
Hamiltonian
\vspace{-0.1cm}
\begin{equation}
\vspace{-0.1cm}
 H = - (\hbar^2/2m) (\partial_x^2 + \partial_y^2) + U \left(x,y \right)
+ V \left( x,y \right) ,
\label{hamiltdisord}
\end{equation}
where $m$ is the electron effective mass, $U$ is the grain boundary potential, and $V$ is the potential due to the edges (see Fig. \ref{Fig-1}
and Refs. \cite{Feilhauer1,Feilhauer2}).
We connect the wire to two ideal leads - clean long wires of width $W$. In the leads, the wave function of the electron with energy $E$ possesses the usual form \cite{Datta-kniha}
\vspace{-0.06cm}
\begin{equation}
\vspace{-0.06cm}
\begin{array}{c}
\varphi(x,y) = {\sum}^{{}_N}_{n=1} \left[ A^+_n(x) + A^-_n(x) \right] \sin(\frac{n \pi y}{W}), \ x\leq 0 \\
\varphi(x,y) = {\sum}^{{}_N}_{n=1} \left[ B^+_n(x) + B^-_n(x) \right] \sin(\frac{n \pi y}{W}), \ x \geq L
\end{array}
\label{rozvoje}
\end{equation}
 where $N$ is the considered number of channels (ideally $N = \infty$), $A^{\pm}_n(x) \equiv a^{\pm}_{n} e^{\pm i k_n x}$, $B^{\pm}_n(x) \equiv b^{\pm}_{n} e^{{\pm} i k_n x}$, and $k_n(E)$ is the wave vector given by equation $\frac{\hbar^2 k^2}{2m}+\frac{\hbar^2 \pi^2 n^2}{2mW^2}=E$.
The vectors $\vc{A}^\pm(0)$ and $\vc{B}^\pm(L)$ with components $A^\pm_{n=1, \dots N}(0)$ and $B^\pm_{n=1, \dots N}(L)$, respectively,
obey the matrix equation
  %
\begin{eqnarray}
\left(
\begin{array}{c}
\vc{A}^-(0) \\
\vc{B}^+(L) \\
\end{array}
\right)
=
\left[
\begin{array}{cc}
r & t' \\
t & r' \\
\end{array}
\right]
\left(
\begin{array}{c}
\vc{A}^+(0) \\
\vc{B}^-(L) \\
\end{array}
\right),
\quad
S
\equiv
\left[
\begin{array}{cc}
r & t' \\
t & r' \\
\end{array}
\right] ,
\label{Smatrixrovnica}
\end{eqnarray}
where $S$ is the scattering matrix \cite{Datta-kniha}. Its elements $t(E)$, $r(E)$, $t'(E)$, and $r'(E)$ are matrices with dimensions $N \times N$. Matrices $t$ and $t'$ are the transmission amplitudes of the waves $\vc{A}^+$ and $\vc{B}^-$, respectively, and matrices $r$ and $r'$ are the corresponding reflection amplitudes. In particular, the matrix element $t_{mn}(E)$ is the amplitude of transmission from channel $n$ in the left lead into the channel $m$ in the right lead. We evaluate $S(E)$ for disorder in figure \ref{Fig-1} following Refs. \cite{Feilhauer1,Feilhauer2}.

\begin{figure}[t!]
\centerline{\includegraphics[clip,width=0.94\columnwidth]{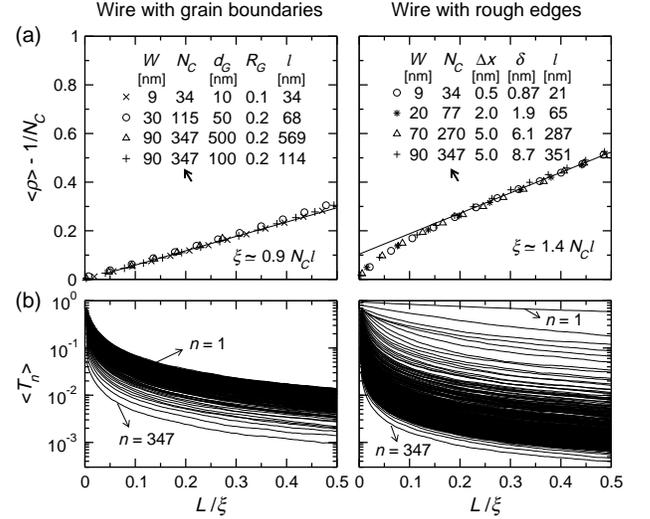}}
\vspace{-0.15cm} \caption{Transport study of the Au wire with grain boundaries [Fig. \ref{Fig-1}(a)] and Au wire with rough edges [Fig. \ref{Fig-1}(b)]. The parameters of Au are $m = 9.1 \times 10^{-31}$kg and $E_F = 5.6$eV, other parameters are listed. Not to affect the results, $N$ is usually kept larger than $N_c$.
Figure (a) shows the mean resistance $\langle \rho \rangle$ versus the wire length $L$.
Note that $\langle \rho \rangle$ is reduced by the contact resistance $1/N_c$ and $L$ scaled by $\xi$. The localization length $\xi$ is obtained from the numerical data for $\langle \ln g \rangle$ (see Refs. [\onlinecite{Feilhauer1,Feilhauer2}]) by means of the fit $\langle \ln g \rangle = - L/\xi$ at $L \gg \xi$. The full lines show the linear fit of the diffusive regime (see the text) from which we obtain the mean free path $l$. In fact,
in the right panel one should see four slightly different full lines for different $N_c$; we show only one of them for simplicity.
 Figure (b) shows $\langle T_n \rangle$ versus $L/\xi$ for the parameters indicated by bold arrows. For $n =1, 2, \dots N_c$ the resulting curves are ordered
 decreasingly.
} \label{Fig-2}
\vspace{-0.5cm}
\end{figure}

The wire conductance $g$ (in units $2e^2/h$) is given by the Landauer formula $g = \sum^{N_c}_{n=1} T_{n}$, where $T_{n}(E_F)  = \sum^{N_c}_{m=1} |t_{mn}(E_F)|^2 \frac{k_m(E_F)}{k_n(E_F)}$ is the transmission probability of channel $n$.
 We evaluate $t_{mn}$ for many samples and obtain the mean transmission $\langle T_n \rangle$ and mean resistance $\langle \rho \rangle = \langle 1/g \rangle$.

 Figure \ref{Fig-2} shows our data. The wires with grain boundaries exhibit features of the white-noise-like disorder. First, $\langle \rho \rangle$ follows the usual diffusive law $\langle \rho \rangle = 1/N_c + (2/k_F l)(L/W)$, shown in a full line. Second, all $\langle T_n \rangle$ are equivalent in the sense that $\langle T_n \rangle \propto 1/L$ \cite{vanRossum}. Therefore, \emph{persistent currents in rings made of such wires should agree with equation \eqref{Igre}}.

 For the wires with rough edges, however, $\langle \rho \rangle$ approaches the diffusive law $\langle \rho \rangle = 1/N^{eff}_c + (2/k_F l)(L/W)$ [the full line in the right panel of Fig. \ref{Fig-2}(a)], where
$1/N^{eff}_c$ is the effective contact resistance due to the $N^{eff}_c$ open channels. For large $N_c$ and small enough $\Delta x$ we find the universal number $N^{eff}_c \simeq 6 \ll N_c$ (see \cite{Feilhauer1} and figure S\ref{Fig-3SM}(b) in \cite{sup}). These open channels dominate also the $\langle T_n \rangle$ data in the right panel of figure \ref{Fig-2}(b). Channel $n=1$ is almost ballistic ($\langle T_1 \rangle \simeq 1$) even for $L = 0.2\xi \simeq 100 l$, when a few other channels with low $n$ show $\langle T_n \rangle$ of the order of $0.1$ and all other channels are diffusive or localized, with $\langle T_n \rangle$ strongly suppressed \cite{Feilhauer1,garcia}.
\emph{What are the persistent currents in rings made of such wires?}

Consider the 2D wire in figure \ref{Fig-1}, but ring-shaped in the plane of the 2D gas and with the wire ends connected. So we have a 2D ring. Since $L \gg W$, we use the cyclic conditions
\vspace{-0.1cm}
\begin{equation}
\vspace{-0.1cm}
\begin{array}{c}
\varphi(0,y) = \exp(- i 2 \pi \Phi/\Phi_0) \varphi(L,y), \\
 \frac{\partial \varphi}{\partial x}(0,y) = \exp(- i 2 \pi \Phi/\Phi_0)  \frac{\partial \varphi}{\partial x}(L,y)
\end{array}
\label{hrpo}
\end{equation}
and follow the standard approach \cite{Cheung} in which the electron states of the ring are described by the Hamiltonian
of the stripe (in our case by Eq. \ref{hamiltdisord})
and by conditions \eqref{hrpo}. In other words, the ring states are assumed to coincide with the stripe states obeying the cyclic conditions \eqref{hrpo}. Say, in the clean ring
$\varphi(x,y) \simeq \sin(\frac{n \pi y}{W}) e^{\pm i k_\nu x}$, where $y$ is the position along the ring radius, $x$-axis  is
bent along the ring circumference, and $k_\nu(E) = \frac{2 \pi}{L}(\nu + \Phi/\Phi_0)$, where $\nu=0,\pm 1,\pm 2, \dots$.

Once the ring states and stripe states coincide (we discuss this assumption in detail below), we can estimate the persistent current in the ring with rough edges intuitively from the transmission of the corresponding stripe (the right panel of figure \ref{Fig-2}b). We do not expect the diffusive result \eqref{Igre} because the feature $\langle T_1 \rangle \simeq 1$ at $L \gg l$ is in contrast with usual diffusive decay $\sim l/L$. Assume roughly that $\langle T_n \rangle = 1$ for $n=1$
and $\langle T_n \rangle \sim l/L$ for all other $n$. In this model, the channel $n=1$ contributes by the ballistic
current $I_{typ} = e v_F/L$ while the total contribution from other channels is diffusive, $I_{typ} \simeq (e v_F/L)(l/L)$, and negligible for $L \gg l$. As a result, \emph{multichannel rings with disorder due to rough edges should support at $L \gg l$ typical
currents $I_{typ} \simeq e v_F/L$.} In other words, the rough edges scatter all electrons except for a small part of those that move (in classical terms) almost in parallel with the edges. This small part, composed mainly of electrons occupying
channel $n=1$, hits the edges rarely and thus moves almost ballistically \cite{sup}. Eventually mainly the electron circulating at the Fermi velocity contributes, as in a one-channel ballistic ring. Thus $I_{typ} \simeq e v_F/L$ albeit $L \gg l$.

   Now we discuss the assumption that the ring states coincide for $L \gg W$ with the stripe states obeying the conditions \eqref{hrpo}.  This \emph{standard approach} describes the ring by the stripe-related Hamiltonian (Eq.\ref{hamiltdisord}) that ignores the ring curvature.
   Consider the clean ring. If the ring curvature is included in the Hamiltonian, it produces the centrifugal force which is $\propto \nu^2$ and which pushes the radial wave functions towards the outer edge of the ring. Consequently, the radial wave functions become localized at the outer edge (especially for large $\nu$) and strongly differ from the form $\sin(\frac{n \pi y}{W})$ even for $L \gg W$. This result is exact in the non-interacting model but fails to describe metallic rings, because the localization of the radial wave functions at the outer edge produces the internal field due to the electron-ion and Hartree-Fock interaction. This internal field, ignored in the non-interacting model, tends to balance the centrifugal force and to delocalize the radial waves throughout the ring cross section. Once the balance is achieved, the resulting radial wave functions have to be close to the stripe-related form $\sin(\frac{n \pi y}{W})$. Just this is implicitly assumed in \emph{the standard approach} that \emph{omits from the Hamiltonian both the ring curvature and the internal field}. In reality a small deviation from $\sin(\frac{n \pi y}{W})$ remains and produces the residual internal field balancing the centrifugal force \cite{Simanek}.

        Consider the \emph{standard approach} in terms of the semiclassical paths. Since the $x$ axis  is
bent along the ring, the $x$-component of any straight-line path in the stripe is bent to follow the ring curvature; this curvature-mediated orbital effect is in fact due to the internal field that balances the centrifugal force. The \emph{standard approach} thus strongly differs from the semiclassical-path-based approach \cite{Jalabert} that includes the ring curvature exactly in the non-interacting model but ignores the internal field. As the internal field is ignored, the paths that govern the wave functions are the straight lines \cite{Jalabert} and the radial wave functions are pushed toward the outer edge (this is manifested by the straight-line paths that hit solely the outer edge \cite{Jalabert}). If $L \gg W$, any straight-line path unavoidably hits the ring edges many times \cite{Jalabert}, unlike our path that circulates almost in parallel with the edges (for further insight see \cite{sup}).

      We now verify our estimates of persistent currents by microscopic calculations that rely on the \emph{standard approach}.
Using Eqs. \eqref{rozvoje}, we write equations \eqref{hrpo} in the matrix form
\vspace{-0.1cm}
\begin{equation}
\vspace{-0.1cm}
\left(
\begin{array}{c}
\vc{A}^-(0) \\
\vc{B}^+(L) \\
\end{array}
\right)
=
\left[
\begin{array}{cc}
0 & Q^{-1}(\phi) \\
Q(\phi) & 0 \\
\end{array}
\right]
\left(
\begin{array}{c}
\vc{A}^+(0) \\
\vc{B}^-(L) \\
\end{array}
\right),
\label{okrajka}
\end{equation}
where $Q$ is the $N \times N$ matrix with terms $Q_{\alpha \beta} = e^{i 2\pi \Phi/\Phi_{0}} \delta_{\alpha \beta}$. Equations \eqref{okrajka} and \eqref{Smatrixrovnica} hold together for discrete energies $E =E_j(\Phi) $ which we find for a given ring
numerically \cite{Feilhauer2}. Then we find $I = - {\sum}_{ \forall E_j \leq E_F} dE_j/d\Phi$ and $I_{typ} \equiv \langle I^2\rangle ^{1/2}$, where $\langle I^2\rangle$ is averaged over a small energy window at $E_F$ \cite{Feilhauer2,sup}.

\begin{figure}[t!]
\centerline{\includegraphics[clip,width=0.94\columnwidth]{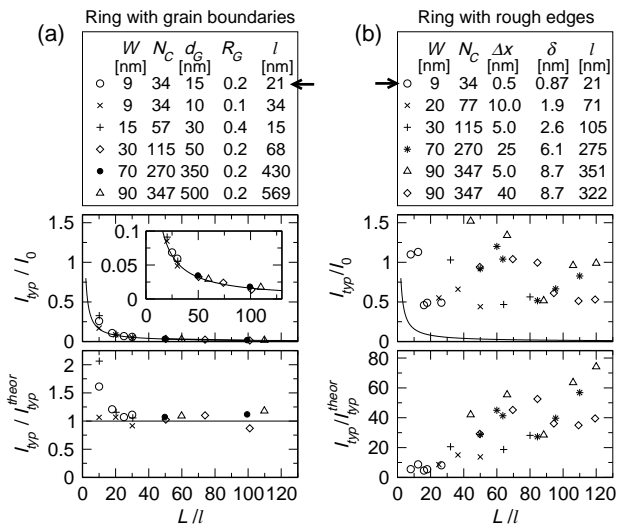}}
\vspace{-0.15cm} \caption{Typical persistent current $I_{typ}$ in a disordered Au ring versus $L/l$. The ring parameters are shown, $\Phi=-0.25 h/e$, $l$ has been obtained from the wire resistivity (figure \ref{Fig-2}). The arrows
point the parameters studied further in Ref. \cite{sup}. Symbols are our data, full lines show the formula $I^{theor}_{typ} = 1.6 (e v_F/L) (l/L)$.} \label{Fig-3}
\vspace{-0.5cm}
\end{figure}

Figure \ref{Fig-3} shows our main results.  \emph{In the rings with grain boundaries, $I_{typ}$ agrees (at large $L$) with expected result $I^{theor}_{typ} = 1.6 (e v_F/L) (l/L)$, like in the experiments \cite{Bluhm,Bles}. However, in
the rings with rough edges, $I_{typ}$ is systematically (not regarding the data fluctuations) close to the ballistic one-channel value $I_0 = e v_F/L$, albeit $L \gg l$, $N_c \gg 1$, and
$\langle \rho \rangle \propto L$. All this agrees with the puzzling experiment \cite{Chand}}.

\begin{figure}[t!]
\centerline{\includegraphics[clip,width=0.85\columnwidth]{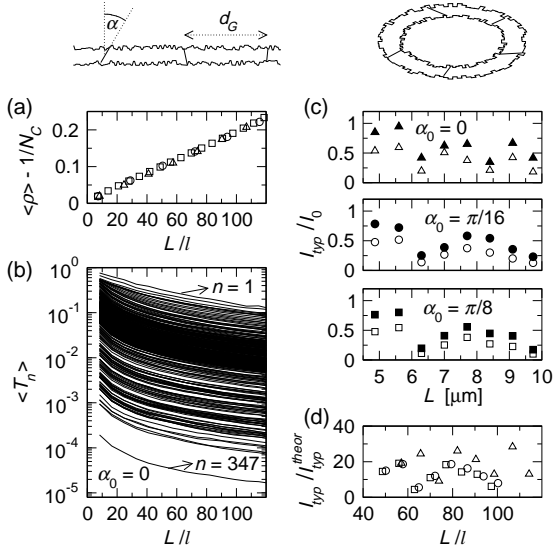}}
\vspace{-0.15cm} \caption{
  Transport in Au wires and Au rings with rough edges and bamboo-like grains. The angle $\alpha$ specifying the orientation of the grain boundary is chosen at random from the interval $(-\alpha_0,\alpha_0)$, where $\alpha_0$ is the parameter: $\alpha_0 = 0$ means the ideal bamboo shape with the boundary perpendicular to the wire \cite{Graham}.
  The table shows all parameters and the resulting $l$ and $\xi$. Figure (a) shows the
   mean resistance $\langle \rho \rangle$ as a function of $L/l$, figure (b) show the transmission $\langle T_n \rangle$ versus $L/l$ for $\alpha_0 = 0$. The open symbols in figure (c) show the typical current in the ring, $I_{typ}/I_0$, as a function of $L$ for various $\alpha_0$, the full symbols show the corresponding maximum currents.
   Figure (d) shows the $I_{typ}$ data from figure (c) normalized by $I^{theor}_{typ} = 1.6 (e v_F/L) (l/L)$ and plotted in dependence on $L/l$.} \label{Fig-6}
\vspace{-0.5cm}
\end{figure}

In the work \cite{Chand}
the persistent current $\sim I_0$ was observed
in the Au ring with $L \simeq 100 l$ and $W = 90$nm.
Indeed, the figure \ref{Fig-3}(b) shows $I_{typ} \sim I_0$ also for $L/l \simeq 100$ and $W = 90$nm. The difference is that in the work \cite{Chand} $l \simeq W$ ($l = 70$nm for $W = 90$nm)
while our values of $l$ in Fig. \ref{Fig-3}(b) [see also Fig. S\ref{Fig-3SM}(a) in \cite{sup}] are
at least two to three times larger than $W$; the edge roughness alone cannot produce $l \simeq W$. In reality
the edge roughness coexists with other types of disorder. Reference \cite{Chand} did not specify disorder in the measured samples, but Webb mentions in Ref. \cite{Kircze} that the grains
in the Au rings of work \cite{Chand} were much larger than $W$ (for instance, in Ref. \cite{WebbWashburn} $d_G \sim 8W$). The grains with $d_G \gg W$ are known as bamboo-like grains \cite{Graham}. Of course, $d_G \gg W$ and $l \simeq W$ \cite{Chand} means $l \ll d_G$, which suggests that the grain boundaries were not the main source of scattering in the work \cite{Chand}.
If the random grain boundaries (or impurities) were the main source of scattering, the
measured persistent current \cite{Chand} would be $\sim (l/L)I_0$ rather than $\sim I_0$ (c.f. Fig. \ref{Fig-3} and Ref. \cite{Feilhauer2}).
What remains is the edge roughness and we have seen that it explains the mysterious coexistence of results $I_{typ} \simeq I_0 $, $L/l \gg 1$, and $\langle \rho \rangle \propto L$.
\emph{What happens if one adds the bamboo-like grains?}

Since $d_G \gg W$, we fit $R_G$ to obtain $l \simeq W$. Figure \ref{Fig-6} shows such a study for the same $W$ and similar $L$ as in the work \cite{Chand}.
In figure \ref{Fig-6}(a) we see again the diffusive law $\langle \rho \rangle \propto L/l$, but now $l \simeq W$, like in the work \cite{Chand}. Figure \ref{Fig-6}(b) shows that the transmission through channels $1$, $2$, and a few more is still large
(between $1$ and $0.1$), though not as large as in the wire with rough edges only [c.f. the right panel of Fig. \ref{Fig-2}(b)]. A suppression of the transmission, caused by a combined effect of the rough edges and bamboo-like grains, is visible for all $347$ channels. Consequently, $l$ is suppressed as well and we have $l \simeq W$.
 Similarly, the typical currents [Figs. \ref{Fig-6}(c) and \ref{Fig-6}(d)] are suppressed compared with the pure edge-roughness case [Fig. \ref{Fig-3}(b)], but they still grossly exceed the law $I^{theor}_{typ} = 1.6 (e v_F/L) (l/L)$.
Figure \ref{Fig-6}(c) shows the maximum currents, because \emph{the work \cite{Chand} in fact reported the current amplitudes rather than $I_{typ}$. These amplitudes were between $\sim 0.1I_0$ and $\sim I_0$ and roughly the same show our data} (full symbols).

In conclusion,
figure  \ref{Fig-3} naturally explains why the experiment \cite{Chand} shows the result $I_{typ} \simeq e v_F/L$ and experiments \cite{Bluhm,Bles} confirm the result $I_{typ} \simeq (e v_F/L) (l/L)$. It suggests that
disorder in samples of works \cite{Bluhm,Bles} was white-noise-like (most likely mainly due to the random grain boundaries), while
the dominant disorder in \cite{Chand} was the edge roughness.

A few remarks at the end. (i) The samples of work \cite{Chand} were 3D, but the 3D effects would change our 2D results insignificantly \cite{sup}. (ii) The step-shaped-roughness model in figure \ref{Fig-1}(b) is universal; our results hold also for any model with a smoothly varying roughness \cite{sup}. (iii) Our results are robust against the change of $N_c$, $\delta$, $\Delta x$, $l$, and $L$ for a broad range
of values. Therefore, the absence of the exact information on the nature of disorder in measured samples \cite{Chand,Bluhm,Bles} is not crucial for our conclusions. Anyway, our values of $\delta$ and $\Delta x$ are realistic
(c.f. figures 1 and 2 in Ref. \cite{Bryan}).
Experiments that would determine
$I_{typ}$ and $l$ in correlation with the parameters of disorder can be useful.
(iv) Note \cite{sup}, that the transmissions $T_{n=1} \simeq 1$ in wires with rough edges have nothing in common with the bimodal distribution $1/\sqrt{(1-T)T^2}$, which exists in any diffusive conductor \cite{vanRossum} and diverges for $T=1$. Transmissions $T$ in the bimodal distribution are the eigen-values of the $t^{+}t$ matrix,
while the meaning of our $T_n$ is different.

We thank the TACC at
The University of Texas at Austin for grid resources. We thank for grant VEGA 2/0206/11.

%

\clearpage

\pagestyle{empty}

\renewcommand{\figurename}{FIG. S}
\setcounter{figure}{0}

\section{Ballistic persistent currents in disordered metallic rings: Origin of puzzling experimental values (supplemental material)}

\begin{center}
 J. Feilhauer and M. Mo\v{s}ko \\
 \vspace{0.25cm}
 \emph{Institute of Electrical Engineering, Slovak Academy of Sciences, 841 04 Bratislava, Slovakia} \\
 \end{center}

 This supplemental material consists of six sections. In section I we explain in detail why our transport results obtained for the roughness model in figure \ref{Fig-1}(b) hold universally also for any other roughness model.
 In section II, the standard (wave-function-based) description of the ring states is compared with the semiclassical-paths-based description.
 In section III we provide a further insight into our result $I_{typ} \simeq e v_F/L$ at $L/l \gg 1$ by showing supplemental numerical data. In section IV we explain why we do not study the mean current and focus solely on the typical current. In section V we explain in detail why our 2D study gives the results that hold very well also for 3D samples. Finally, in section VI we stress that the transmission $T_{n=1} \simeq 1$ in the wire with rough edges (which is responsible for the result $I_{typ} \simeq e v_F/L$ at $L/l \gg 1$) should not be confused with a well-known general property of any diffusive conductor, with the transmission eigen-value $T=1$.

 \vspace{-0.3cm}
\subsection{I. On the universality of the roughness model in figure \ref{Fig-1}(b)}

All our transport results for the quasi-1D systems with the rough edges, presented in the main text, were obtained for the step-shaped-roughness model in figure \ref{Fig-1}(b). Here we wish to point out again that these transport results are universal in the sense that they would remain the same also for any model with a smoothly varying roughness.

Evidently, from the technical point of view, the step-shaped roughness in figure \ref{Fig-1}(b) provides a discretization scheme that allows to model any smoothly varying roughness by means of the very small and very dense steps. Using this approach, our calculations from the main text can in principle be repeated for any roughness model which is specified by the RMS parameter $\delta$ and roughness correlation length $\Delta x$. We expect that the obtained transport results will agree with the results presented in the main text, if one compares the dependencies on the parameter $L/\xi$, where $\xi$ is the localization length. This expectation is motivated by a few fundamental findings.

First, the statistical ensemble of the macroscopically-identical mesoscopic conductors with impurity disorder is known to exhibit the conductance distribution which is essentially the same (for a given value of parameter $L/\xi$) for any model of the impurity disorder model; the weaker the disorder the better the accord of the conductance distribution of various models). Second, it seems that a similar universality (the independence on the specific model of disorder) holds also when the impurity disorder is replaced by disorder due to the edge roughness. In particular, the conductance calculations in Ref. \cite{garcia}, performed for the same step-shaped-roughness model as our model in figure \ref{Fig-1}(b), gives a quite similar results as the conductance calculations in the paper [SM1], performed for the smoothly varying roughness with the Gaussian correlation function. To demonstrate this universality by means of the direct comparison, we have performed the conductance calculations for the smoothly-varying roughness with the Gaussian correlation function (the model of Ref. [SM1], and we have compared them with our results for the step-shaped roughness in figure \ref{Fig-1}(b).

\begin{figure}[t!]
\centerline{\includegraphics[clip,width=1.0\columnwidth]{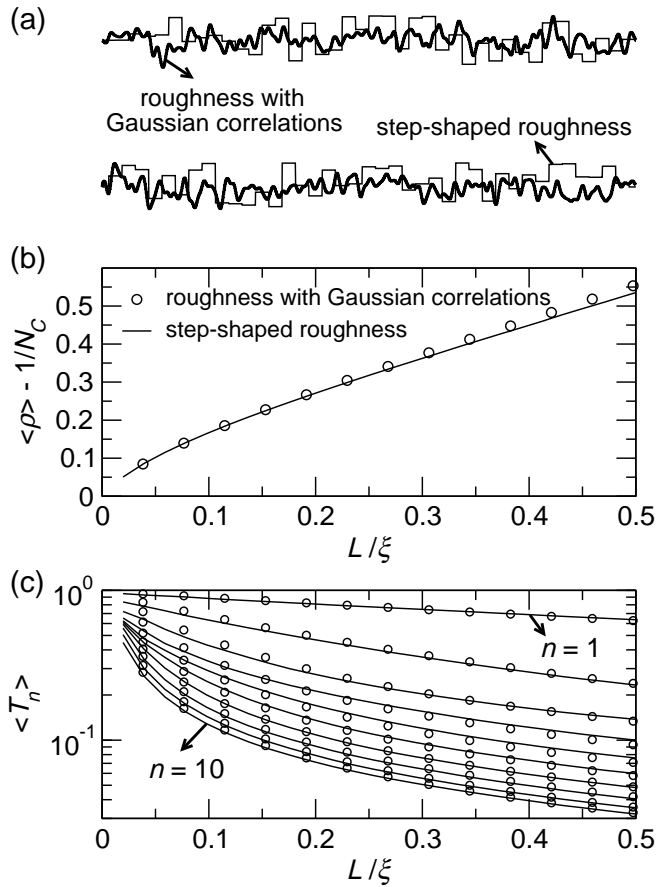}}
\vspace{-0.15cm} \caption{(a) The top view on the 2D wire with the rough edges generated numerically for two different roughness models.
In this numerical example the Au wire of width $W = 9$nm is considered, which implies that the number of the conducting channels ($N_c$) is $34$. The RMS roughness amplitude $\delta$ and roughness-correlation length $\Delta x$
are $\delta = 0.87$nm and $\Delta x = 0.5$nm for the step-shaped roughness, for the roughness with the Gaussian correlations $\delta = 0.5$nm and $\Delta x = 1.2$nm. Using the approach described in the main text, in the former case we obtain the mean free path $l = 21$nm and localization length $\xi \simeq 1.4 N_c l$, and in the latter case we get $l = 20.8$nm and $\xi \simeq 1.49 N_c l$. (b) The mean resistance $\langle \rho \rangle$ as a function of the dimensionless wire length $L/\xi$;
a comparison for the roughness models specified above. (c) The same comparative study as in figure (b), but for the mean channel transmissions $\langle T_n \rangle$; for clarity only the data for the first ten conducting channels are presented.
} \label{Fig-3SM}
\vspace{-0.5cm}
\end{figure}

In figure \ref{Fig-3SM} we show a typical output of such comparative study for two Au wires with the same number of the conducting channels ($N_c = 34$), so that one can compare directly the individual channel transmission. It can be seen that the individual transmissions are in a very good agreement, which illustrates the above mentioned universality; note also that the individual transmissions for both roughness models coincide albeit the values of the parameters $\delta$ and $\Delta x$ in the considered roughness models are (intentionally) not the same. In addition, the universality with respect to the choice of $\delta$ and $\Delta x$ within the same roughness model is obvious for all our data in the main text and especially from our paper \cite{Feilhauer1}.
The last but not least, our main result (the ballistic-like persistent current $I_{typ} \simeq ev_F/L$ in figure \ref{Fig-3}(b)) is universal simply because of the absence of the sensitivity to the edge roughness.

Any user of the scattering-matrix technique can perform a similar universality demonstration for other types of the smooth roughness. The prize for the use of the smoothly varying roughness is a much longer computational time, which is crucial especially for the persistent current calculations. Due to this reason our study in the main text relies on the step-shaped roughness, however, our results are universal as mentioned above.

\vspace{-0.3cm}
\subsection{II. Comparison of the standard wave-function-based approach with the semiclassical-path-based approach}

The persistent current calculations presented in the main text rely on the \emph{standard approach}.
The key assumption of the standard approach, justified in the main text, is that the ring states coincide for $L \gg W$ with the stripe states obeying the conditions \eqref{hrpo}. Just this assumption allows to describe the ring states by the stripe-related Hamiltonian (Eq.\ref{hamiltdisord}) that ignores the ring curvature. The main text also mentions that the \emph{standard approach} strongly differs from the semiclassical-path-based approach \cite{Jalabert} that incorporates the ring curvature rigorously (within the non-interacting model) but ignores the internal Hartree-Fock-interaction-mediated field balancing the centrifugal force. In this approach the semiclassical paths in the ring fundamentally differ from those in the stripe, or in other words, the ring states and stripe states do not coincide. We wish to give a few more comments on the semiclassical-path-based approach \cite{Jalabert}.

Consider first the clean ring. In spite of the fact that the radial wave functions are pushed towards the outer edge of the ring by the centrifugal force, the semiclassical-path-based approach by Jalabert et al. \cite{Jalabert} gives correct result for the ballistic persistent current.
The problem arises when one considers the rough edges (see Samohkin's paper \cite{Jalabert}). Since only the straight-line paths are allowed (neglecting the orbital effect due to the magnetic field), in the annular geometry with $L \gg W$ any straight-line path unavoidably hits the ring edges many times when it makes one trip around the ring \cite{Jalabert}. On the contrary, the stripe geometry allows also very long straight-line paths that are (almost) parallel with the stripe edges and therefore do not feel the edge roughness. Within the approach of Refs. \cite{Jalabert}, these very long ballistic paths are changed in the annular geometry on the paths that hit exclusively the outer edge of the ring and become scattered by the edge roughness.
On the other hand, in our \emph{standard approach} these paths remain parallel with the ring edges and carry the ballistic-like current. This fundamental difference is due to the fact that the \emph{standard approach} implicitly incorporates the balance between the centrifugal force and internal Hartree-Fock field (see the main text), while in the semiclassical-path-based approach \cite{Jalabert} the centrifugal force remains unbalanced.

\begin{figure}[b!]
\centerline{\includegraphics[clip,width=1.0\columnwidth]{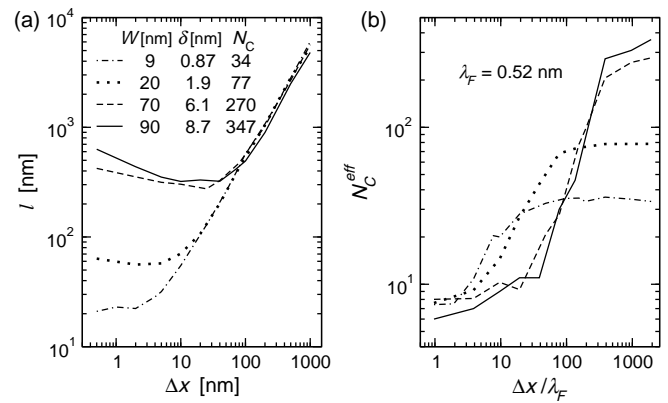}}
\vspace{-0.15cm} \caption{(a) The mean free path $l$ and (b) effective number of the open channels $N^{eff}_c$ in the wire with rough edges, both plotted in dependence on the roughness correlation length $\Delta x$ for the parameters as indicated. These data were extracted from the data for
$\langle \rho \rangle$ versus $L$ by means of the fit $\langle \rho \rangle = 1/N^{eff}_c + (2/k_F l)(L/W)$ as is explained in Fig. \ref{Fig-2}(a) and in the main text.
For simplicity, $\delta/W$
is kept nearly the same ($\sim 1/10$) for each set of $\delta$ and $W$.} \label{Fig-3SM}
\vspace{-0.5cm}
\end{figure}

We also wish to address another fundamental difference. Samokhin \cite{Jalabert} assumed that the straight-line path that hits the ring edge is reflected diffusively no matter what is the incidence angle (the angle between the path and the edge). Specifically, the probability of the diffusive reflection from the edge is described by the Fuchs coefficient which is equal to unity for all incidence angles. It should be mentioned that a realistic probability of the diffusive reflection, derived by Soffer and Ziman [SM2] for
a free wave impinging the surface with uncorrelated roughness, strongly depends on the incidence angle. In particular, it is equal to unity
for perpendicular incidence but approaches zero for small incidence angles. Samokhin \cite{Jalabert} found in the ring with rough edges the diffusive persistent current. However, he would certainly not find such diffusive current, if the Fuchs coefficient is replaced by the angle-dependent probability of the diffusive reflection due to Ziman and Soffer [SM2]: Owing to the (almost) specular reflections at small angles, his result would
most likely become more similar to our ballistic prediction.
In our paper the correct angle dependence of the edge roughness scattering is included microscopically in the scattering matrix method.

Indeed, the tendency to a specular reflection at small angles is manifested by the channel transmission $T_n$. Let us look at the right panel of figure 2b in detail.
Semiclassically, the channel number $n$ corresponds to the angle between the semiclassical trajectory and edge, and $n=1$ corresponds to the smallest nonzero
semiclassical angle allowed by the quantum confinement. Consider, say $L \simeq 0.25 \xi \simeq 120 l$. In case of the diffusive reflection, for $L/l = 120$ one should observe $T_n \sim l/L \sim 1/120$ already for $n=1$. However, this is not the case; the right panel of figure 2b  shows that $T_n$ is between 1 and 0.1 for $n =1, 2, \dots, 6$. In semiclassical terminology, transmissions for "angles" $n =1, 2, \dots, 6$ exceed the diffusive value $\sim l/L \sim 1/120$ by one to two orders of magnitude. This means that the motion within these channels is much more ballistic than diffusive, although (semiclassically speaking) the electrons in these channels experience multiple collisions with the edges. The diffusive values of $T_n$ are reached only if $n$ is large enough, that means for the large enough semiclassical angles.

\subsection{III. Supplemental numerical data}
\vspace{-0.5cm}
The figure S\ref{Fig-3SM} shows the electron mean free path $l$ and the effective number of the open channels $N^{eff}_c$ in the wires with rough edges, extracted from the linear fit of the numerical data for $\langle \rho \rangle$ versus $L$ (see the discussion of figure \ref{Fig-2} in the main text). Two features in the figure are worth noticing.

First, the $l$ versus  $\Delta x$ dependence in figure S\ref{Fig-3SM}(a) demonstrates clearly that the minimum mean free path due to the edge roughness scattering is always a few times larger than the wire width $W$. In other words, the edge roughness alone cannot explain the observation $l \simeq W$, reported in the experiment \cite{Chand}. To explain the observed persistent currents $I \simeq e v_F/L$ jointly with observation $l \simeq w$, the edge roughness has to be combined with the bamboo-like grains, as is demonstrated in figure \ref{Fig-6} in the main text.

Second, the $N^{eff}_c$ versus $\Delta x$ dependence in figure S\ref{Fig-3SM}(b) demonstrates  that $N^{eff}_c$ is a universal ($N_c$-independent) number of the order of $10$ if $\Delta x$ is small enough and $N_c$ large. The universal $N^{eff}_c$  has been discovered in Ref. \cite{Feilhauer1}, here the universality is demonstrated for $N_c$ as large as $347$. Of course, the rings made of such wires have to possess the same $N^{eff}_c$. Now we demonstrate that this is indeed the case.

\begin{figure}[t!]
\centerline{\includegraphics[clip,width=1.0\columnwidth]{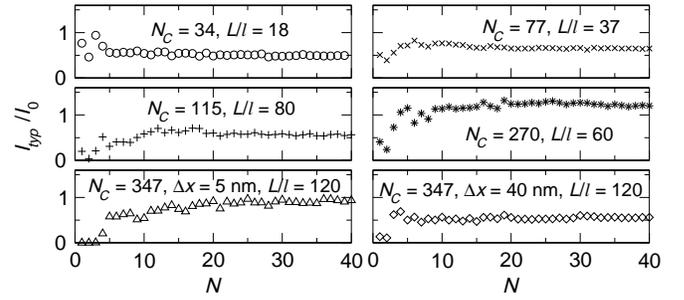}}
\vspace{-0.15cm} \caption{Typical persistent current $I_{typ}$ in the ring with rough edges as a function of the total number of channels ($N$) considered in the simulation. The same parameters and symbols are used as
in figure \ref{Fig-3}(b), the considered ring lengths are shown as $L/l$.} \label{Fig-4}
\vspace{-0.5cm}
\end{figure}

In figure S\ref{Fig-4} we show how the typical current in the ring with rough edges depends on the number of
channels ($N$) considered in the simulation. It is (roughly) $N$-independent for $N \gtrsim 10$, no matter how large $N_c$ is. In other words, the currents $\sim I_0$ in rings with rough edges are almost exclusively carried by the open channels $n = 1, 2, \dots, N^{eff}_c$, where $N^{eff}_c \sim 10$ for any value of $N_c$.

 Moreover, as the transmission of these open channels is large for $L/l \gg 1$ [in particular, $\langle T_1 \rangle \sim 1$, as is shown in the right panel of figure \ref{Fig-2}(b)], one also understands why the ring with rough edges supports the ballistic-like current $I_{typ} \sim I_0$ rather than the diffusive current $I_{typ} \sim (l/L) I_0$.
 The current $I_{typ} \sim I_0$ exists because the rough edges scatter all electrons except for a small part of those that move (classically speaking) almost in parallel with the edges. This small part, composed mainly of electrons occupying
channel $n=1$, hits the rough edges rarely and therefore moves almost ballistically. The resulting current, $I_{typ} \sim e v_F/L$, is due the electron that circulates with Fermi velocity, because contributions from other electrons tend to cancel like in a true single-channel ballistic ring.

 Since $\langle T_1 \rangle \sim 1$, one could naively think that the value $I_{typ} \sim I_0$ will survive also if one chooses $N$ as small as $N=1$. Figure S\ref{Fig-4} shows that this is not the case. For instance, in the ring with $N_c = 347$ and $L/l = 120$ the current approaches zero just for $N \rightarrow 1$. This is easy to understand: Once the channel $n = 1$ cannot communicate with other channels,
 the transmission $\langle T_1 \rangle \sim 1$ tends to be suppressed to zero by Anderson localization, present in any sufficiently long 1D disordered system. Communication with a few other channels is needed to restore $\langle T_1 \rangle \sim 1$ and to obtain $I_{typ} \sim I_0$.

\begin{figure}[t!]
\centerline{\includegraphics[clip,width=1.0\columnwidth]{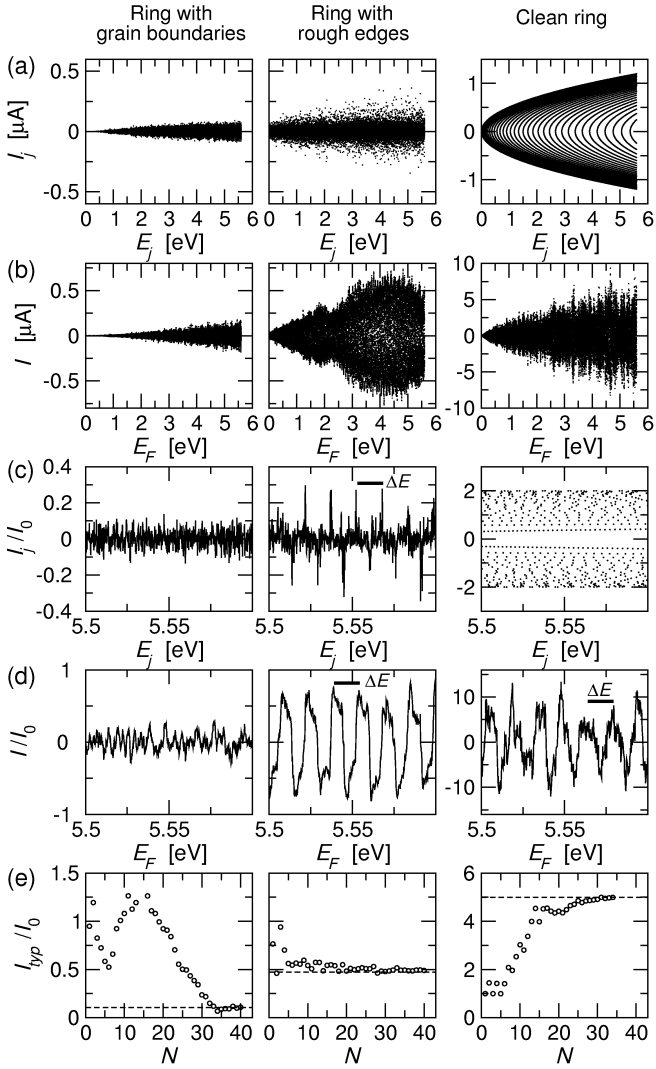}}
\vspace{-0.15cm} \caption{Persistent currents in a ring with grain boundaries, a ring with rough edges, and a clean ring for the parameters marked by the arrows in Fig.~\ref{Fig-3}, for $L=375$ nm, and for $\Phi=-0.25 h/e$. For both disordered rings, the considered parameters ensure
$l(E_F)=21$ nm at the Au Fermi level ($E_F = 5.6$ eV). Figure (a) shows the single-electron current $I_j$ versus the eigen-energy $E_j$. Figure (b) shows the total current $I = \sum_{ \forall E_j \leq E_F} I_j$ obtained by summing the currents in the figure (a) for
$E_F$ varied from $0$ to $5.6$ eV. Figures (c) and (d) show the same data as the figures (a) and (b), but for a small energy window below the Au Fermi level. The data are scaled by $I_0= e v_F/L$,
the data points are connected by full lines which serve as a guide for the eye, the bars depict the energy increment $\Delta E = 2 \pi \hbar v_F /L$.
Figure (e) shows the typical current $I_{typ} \equiv \langle I^2\rangle ^{1/2}$.
Averaging over the energy window in figure (d) gives the values shown by dashed lines:
$I_{typ}/I_0 \simeq 1.6 (l/L)$ for the ring with grain boundaries, $I_{typ}/I_0 \simeq 0.5$ for
the ring with rough edges, and $I_{typ}/I_0 \simeq \sqrt{N_c}$
for the clean ring. The circles show the data obtained by varying
the number of channels, $N$, from $N=1$ to $N > N_c$ (here $N_c = 34$). } \label{Fig-7}
\vspace{-0.5cm}
\end{figure}

To provide further insight, figure S\ref{Fig-7} shows the sample-specific currents in two selected rings from figure \ref{Fig-3} (bold arrows) and in a clean ring. Figure S\ref{Fig-7}(a) shows the dependence $I_j$ versus $E_j$, figure S\ref{Fig-7}(b) shows the total current $I = \sum_{ \forall E_j \leq E_F} I_j$ versus $E_F$.
 Evidently, the ring with rough edges exhibits remarkably larger currents than the ring with grain boundaries, albeit both rings are of the same size and posses the same value of $l$.

 Figures S\ref{Fig-7}(c) and S\ref{Fig-7}(d) focus on a small energy window below the Au Fermi level. One can see that $I_j$ in the ring with rough edges exhibits sharp peaks with the sign alternating and oscillating with period
 $\Delta E = 2 \pi \hbar v_F /L$. This period is twice the inter-level distance in the ballistic single-channel ring,
 which suggests that the peaks are due to the quasi-ballistic channel $n=1$. [We recall that $\langle T_1 \rangle \sim 1$ also for $L/l \gg 1$, as is shown in the right panel of figure \ref{Fig-2}(b).] However, the height of the peaks is affected also by other channels, because, as discussed above, channel $1$ cannot keep $\langle T_1 \rangle \sim 1$ without communicating with a few other channels.

 In figure S\ref{Fig-7}(d) one can see that in the ring with rough edges
 also the total current $I(E_F)$
 oscillates with period $\Delta E$. The amplitudes of the total current are close to $I_0$, and therefore the typical currents
 of size $\sim I_0$ appear in figure \ref{Fig-3}(b).

In fact, already the data for the clean ring
show $I(E_F)$ oscillating with period
$\Delta E$. However, the amplitude of $I$ is $\sim \sqrt{N_c}2I_0$ and the amplitude of $I_n$ is $2I_0$, where
the factor of $2$ is due to the spin.
Evidently, the rough edges reduce $I$ from $\sim \sqrt{N_c}2I_0$ to $\sim I_0$, but they do not change the oscillation period set by the clean ring. Note that also the ring with grain boundaries exhibits the oscillating persistent current. These oscillations are chaotic and correlated with correlation length $\sim (l/L) \Delta E$, predicted \cite{Cheung} for the white-noise-like disorder.

Figure S\ref{Fig-7}(e) shows the typical current.
The dashed lines show the values of $I_{typ}$ obtained from the data in figure S\ref{Fig-7}(d), the circles
show $I_{typ}$ in dependence on $N$.  For all three rings one sees, that the circles approach with raising $N$ the $N$-independent value (the large $N$ limit)
represented by the dashed line. It can be seen that a reliable estimate of $I_{typ}$ in the ring with grain boundaries requires $N \gtrsim N_c$, while for the ring with rough edges one only needs $N \sim 10$ no matter how large $N_c$ is. This is due to the effective number $N^{eff}_c \sim 10$, as
has already been explained in the beginning of this section.

\subsection{IV. The problem of the mean persistent current}

In our present work we have focused on the typical current and we have not discussed the mean current. The sign and amplitude of the mean current, measured in the pioneer experiment by Levy et al. \cite{Buttiker} is another puzzling problem in the field. As can be seen from our description of the scattering matrix method, we can in principle provide also numerical data for the average persistent current. There are however a few serious reasons why our present manuscript is not focused on the average current.

First, the problem of the average current has been addressed by Bary-Soroker et al. [SM3] who attempted to explain it within the interacting electron model. However, these authors did not address the problem of the typical current, and the experiments \cite{Bluhm,Bles} showed, that the typical current is most likely not affected by electron-electron interaction and should be tractable within the non-interacting model. Just these reasons lead us to focus on the problem of the giant typical current and to solve it within the single-electron model.

Second, a complete scattering-matrix study of the amplitude and sign of the average current would require (perhaps) ten to hundred times more computational time than the study of the typical current, presented in our present manuscript. We have therefore decided to focus on the problem of the typical current, and already this problem was computationally cost.

\subsection{V. On the robustness of our 2D results against the 3D effects}

All our transport data in the main text were obtained within the 2D model depicted in figure \ref{Fig-1}, while the experimental samples
of reference \cite{Chand} were three-dimensional. Here we want to point out in detail that the extension of our 2D study to 3D (replacement of the rough edges by rough side-walls) would not change our major results remarkably. The effect of 3D on our 2D results can be estimated easy without any explicit calculation as follows.

In our 2D wire (\ref{Fig-1} (b)) only the edge roughness scattering is considered, while in the 3D wire of reference \cite{Chand} the roughness scattering is in general due to the wire edges (side walls) and due to the top and bottom surfaces in addition. In spite of this difference it is evident that the 3D sample still preserves the key feature of our 2D model: The electrons which occupy the ground 1D channel (now the channel with quantum numbers $n_y = 1$ and $n_z = 1$, where $z$ is the vertical direction) still move almost in parallel with the sample edges and sample surfaces, and therefore avoid the roughness scattering quite similarly as in the 2D case. Due to this key feature, the almost perfect transmission of the Fermi electron in the ground 1D channel, found for the 2D wire (Fig.2), has to persist also in the 3D wire. Owing to this ballistic Fermi electron, the 3D ring has to carry the persistent current $I_{typ} \simeq ev_F/L$ also for $L/l >> 1$, just like the 2D rings in figure \ref{Fig-3}(b).

Furthermore, one can easy see that the roughness-mediated scattering in 3D does not modifies the mean free path $l$ in comparison with 2D remarkably. As already mentioned, in the 3D wires one should consider also the roughness-mediated scattering from the top surface and bottom surface. However, in real normal-metal 3D wires the roughness amplitude (RMS) of the top and bottom surfaces is usually of the order of one lattice constant ($\sim 0.5$nm; see e.g. the paper [SM4]),  which is one order of magnitude less than the roughness amplitude at the edges (RMS $\sim 5$nm - $10$nm; see the experiment of \cite{Bryan}, and our manuscript). Since the roughness-limited mean free path is proportional to the square of the RMS (see e.g. our paper \cite{Feilhauer1} or references … cited therein), the effect of the top and bottom surfaces on the mean free path is two orders of magnitude weaker than the effect of the edges. In other words, it is very likely, that in the 3D wires of reference \cite{Chand} the roughness-mediated scattering is mainly due to the wire edges, while the reflections from the almost smooth top/bottom surface are (almost) specular and thus do not affect the electron transport. The case when the roughness of the top and bottom surfaces is comparable with the edge roughness is experimentally unlike: The roughness of the edges is due to the limitations of the electron lithography and lift-off, due to the roughness of the resist side walls, etc. (see \cite{Bryan}), while the origin of the surface roughness is quite different. However, even in this unlike case the rough surfaces would marginally affect the quasi-ballistic channel $n_y = 1$ and $n_z = 1$ and they would only reduce the edge-roughness-limited mean free path by a factor of $\sim 2$, as can be estimated from the Mathiessen rule.

Finally, unlike our 2D wire in figure \ref{Fig-1} (b), the edges of the 3D wire are in fact the side walls and the edge roughness at such side walls in general scatters the electrons also into the vertical ($z$) direction, in addition to the in-plane scattering considered in our 2D model. This may decrease the roughness-limited mean free path say by a few tens of percent, but cannot affect the fundamental feature (the ballistic-like motion in the ground 1D channel) responsible for the persistent current $I_{typ} \simeq ev_F/L$ also for $L/l >> 1$.

 In principle, our scattering-matrix approach allows to produce also the numerical data for the 3D samples (we already did so for the wires with the grain boundaries
 in reference \cite{Feilhauer2}), albeit not for the samples as large as those used in the experiment \cite{Chand}. However, the computational time would be quite huge and we do not expect any new features in comparison with our 2D results.

 \subsection{VI. Feature $T_{n=1} \simeq 1$ in the wire with rough edges and $T=1$ as a general feature of any diffusive wire: Two different things}

  We recall that the ring with the rough edges exhibits the persistent current $I_{typ} \simeq ev_F/L$ at $L \gg l$ (figure \ref{Fig-3}(b)), because the constituting rough wire exhibits the transmission $T_{n=1} \simeq 1$ (right panel of figure \ref{Fig-2}(b)). We want to stress that the transmission $T_{n=1} \simeq 1$ in the wire with rough edges has nothing in common with the well-known bimodal distribution $1/\sqrt{(1-T)T^2}$, which exist in any diffusive conductor \cite{vanRossum} and diverges for $T=1$. Note that the transmissions $T$ in the bimodal distribution are the eigen-values of the $t^{+}t$ matrix \cite{vanRossum},
while we speak about $T_n=\sum_{m=1}^{N_C}|t_{n,m}|^2$, which are the diagonal elements of the $t^{+}t$ matrix.
 In other words, the channels corresponding to the eigen-values $T$ in the distribution $1/\sqrt{(1-T)T^2}$ are the eigen-states of the $t^{+}t$ matrix, while the channels corresponding to our diagonal elements $T_n$ are the usual plane-wave states. To avoid misunderstanding by readers who are accustomed to the latter definition, we add a few more remarks.

The bimodal distribution $1/\sqrt{(1-T)T^2}$ as a general property of any diffusive conductor with white-noise-like disorder \cite{vanRossum} coexists with the diffusive persistent current $I_{typ} \simeq (ev_F/L)(l/L)$ in the corresponding disordered ring \cite{Cheung}. In other words, the eigenvalues $T=1$ in the bimodal distribution do not cause any ballistic persistent current. The reason why the current is diffusive in spite of $T=1$, is most likely that the eigenvalue $T=1$ does not necessarily mean the ballistic motion (a well known example is the perfect transmission in case of resonant tunneling). The situation becomes fundamentally different for disorder due to the rough edges. In this case the eigen-values $T_n$ still follow the bimodal distribution $1/\sqrt{(1-T)T^2}$, however, this has nothing in common with the ballistic-like persistent current found by us. The ballistic-like current is due to the appearance of the diagonal element $T_{n=1} \simeq 1$: Namely, any wire from the statistical ensemble of wires with rough edges exhibits the diagonal element $T_{n=1} \simeq 1$ independently on the choice of the Fermi energy and wire length. These features are the attributes of ballistic electron motion within channel $n=1$. Indeed, it is easy to check in our simulation, that the electron plane wave entering the wire in the channel $n=1$ remains (almost) unscattered between any two successive scatterers inside the disordered region. As a result, the ring made of such wire supports the persistent current with typical size dominated by the ballistic channel $n = 1$, that is, $I_{typ} \simeq ev_F/L$. In summary, the reason for appearance of $I_{typ} \simeq ev_F/L$ is the ballistic behavior of the diagonal element $T_{n=1}$; the fact that the bimodal distribution gives eigenvalues $T=1$ is irrelevant.


\vspace{0.4cm}
[SM1] J. A. S\'anchez-Gil, V. Freilikher, A. A. Maradudin, and I. V. Yurkevich, Phys. Rev. B \textbf{59}, 5915 (1999).

[SM2] J. M. Ziman, \emph{Electrons and Phonons} (Oxford University Press, London, UK, 1960); S. B. Soffer, \emph{J. Appl. Phys.} \textbf{38}, 1710 (1967).

[SM3] H. Bary-Soroker, O. Entin-Wohlman, Y. Imry, \emph{Phys. Rev. Lett.} \textbf{101}, 057001 (2008).

[SM4] R. C. Munoz, G. Vidal, G. Kremer, L. Moraga, C. Arenas, and A. Concha, \emph{J. Phys.: Conference Series}  \textbf{12}, 2903 (2000).

\end{document}